\documentclass[aps,pre,preprint,preprintnumbers,amsmath,amssymb,a4paper]{revtex4-1}

\usepackage{url}
\date{}

\usepackage{fancyhdr,graphicx}
\usepackage{epstopdf}

\begin{document}

\title{Role of the Number of Microtubules in Chromosome Segregation during Cell Division}

\author{Zsolt Bertalan}
\affiliation{ISI Foundation, Via Alassio 11/c, 10126 Torino, Italy}
\author{Zoe Budrikis}
\affiliation{ISI Foundation, Via Alassio 11/c, 10126 Torino, Italy}
\author{Caterina A. M. La Porta}
\email{caterina.laporta@unimi.it}
\affiliation{Center for Complexity and Biosystems, Department of Bioscience, University of Milan, via Celoria 26, 20133 Milano, Italy}
\author{Stefano Zapperi}
\email{stefano.zapperi@unimi.it}
\affiliation{ISI Foundation, Via Alassio 11/c, 10126 Torino, Italy}
\affiliation{Center for Complexity and Biosystems, Department of Physics, University of Milan, via Celoria 16, 20133 Milano, Italy}
\affiliation{CNR-IENI, Via R. Cozzi 53, 20125 Milano, Italy}
\affiliation{Department of Applied Physics, Aalto University, P.O. Box 14100, FIN-00076 Aalto, Espoo, Finland}

\begin{abstract}
Faithful segregation of genetic material during cell division requires alignment of chromosomes between two spindle poles and attachment of their kinetochores to each of the poles. Failure of these complex dynamical processes leads to chromosomal instability (CIN), a characteristic feature of several diseases including cancer.  While a multitude of biological factors regulating chromosome congression and bi-orientation have been identified, it is still unclear how they are integrated so that coherent chromosome motion emerges from a large collection of random and deterministic processes. Here we address this issue by a three dimensional computational model of motor-driven chromosome congression and bi-orientation during mitosis. Our model reveals that successful cell division requires control of the total number of microtubules: if this number is too small bi-orientation fails, while if it is too large not all the chromosomes are able to congress. The optimal number of microtubules predicted by our model compares well with early observations in mammalian cell spindles. Our results shed new light on the origin of several pathological conditions related to chromosomal instability.
\end{abstract}

\maketitle

\section*{Introduction}
Cell division is a complex biological process whose success crucially depends on the 
correct segregation of the genetic material enclosed in chromosomes into the two daughter cells.  
Successful division requires that chromosomes should align on a central plate between the two poles
of an extensive microtubule (MT) structure, called the mitotic spindle, in a process known as congression \cite{magidson2011}. Furthermore, the central region of each chromosome, the kinetochore, should attach to MTs emanating from each of the two poles, a condition known as bi-orientation \cite{walczak2010}. Only when this arrangement is reached, do chromosomes split into two chromatid sisters that are then synchronously transported towards the poles \cite{matos2009}. Failure for chromosomes to congress or bi-orient can induce mitotic errors which lead to chromosomal instability (CIN), a state of altered chromosome number, also known as aneuploidy. CIN is a characteristic feature of human solid  tumors and of many hematological malignancies \cite{boveri1903}, a principal contributor to genetic heterogeneity in cancer \cite{burrell2013} and an important determinant of clinical prognosis and therapeutic resistance \cite{lee2011,bakhoum2012}. 

\begin{figure}[p]
	\begin{center}
		\includegraphics[width=0.4\columnwidth]{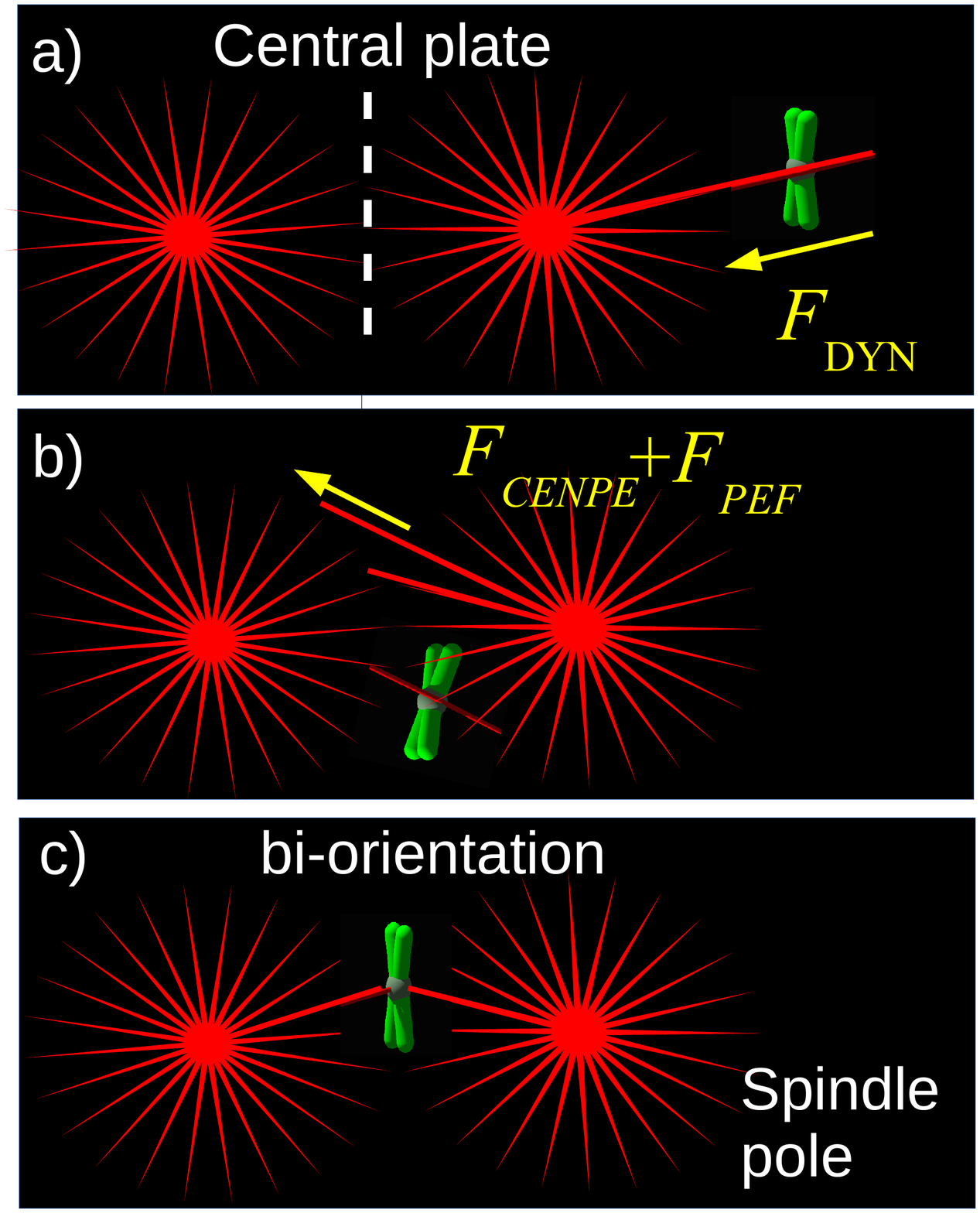}
	\end{center}
	\caption{Schematic of the dynamics of a single chromosome. a) Peripheral chromosomes, not
		lying between the spindle poles, are driven to the nearest pole by dynein. b) Chromosomes are
		driven from the pole to the central plate by the combined action of CENP-E and PEF. c) At the central plate,
		chromosomes attached to both poles are called bi-oriented.}
	\label{fig:newcartoon}
\end{figure}

Chromosome congression occurs in a rapidly fluctuating environment since the mitotic spindle is constantly changing due to random MT polymerization and depolymerization events. This process, known as dynamic instability, 
is thought to provide a simple mechanism for MTs to search-and-capture all the chromosomes scattered throughout the cell after nuclear envelope breakdown (NEB)\cite{kirschner1986}. 
Once chromosomes are captured, they are transported to the central plate by molecular motors that use MTs as tracks. The main motor proteins implicated in this process are kinetochore dynein,
which moves towards the spindle pole (i.e. the MT minus end) \cite{rieder1990,li2007,yang2007,vorozhko2008} and  centromere protein E  (CENP-E or kinesin-7) \cite{kapoor2006,cai2009,barisic2014} and polar ejection forces (PEFs) \cite{Rieder1986}, both moving away from the pole (i.e. they are directed towards the MT plus end). PEFs mainly originate from kinesin-10 (Kid) and are antagonized by kinesin-4 (Kif4A) motors \cite{stumpff2012}, sitting on chromosome arms\cite{barisic2012}. While PEFs are not necessary for chromosome congression, they are vital for cell division \cite{barisic2014} since they orient chromosome arms \cite{barisic2012}, indirectly stabilize end-on attached MTs \cite{cane2013} and are even able to align chromosomes in the absence of kinetochores \cite{cai2009}. Recent experimental results show that chromosome transport is first driven towards the poles by dynein and later towards the center of the cell by CENP-E and PEF \cite{barisic2014} (see Fig \ref{fig:newcartoon}). 

A quantitative understanding of chromosome congression has been the goal of intense theoretical research 
focusing on the mechanisms for chromosome search-and-capture \cite{holy-leiber-1995,wollman2005,paul2009},  motor driven dynamics \cite{joglekar2002,civelekoglu2006,gardner2005,chacon2013,gluncic2015} and attachments
with MTs \cite{hill1985,bertalan2014}.  A mathematical study of search-and-capture was
performed by Holy and Leibler who computed the rate for a single MT to find a chromosome by randomly exploring a spherical region around the pole \cite{holy-leiber-1995}. Later, however, Wollman et al. \cite{wollman2005} showed numerically that a few hundred MTs would take about an hour to search and capture a chromosome, instead of few minutes  as observed experimentally. It was therefore argued that MTs should be chemically biased towards the chromosomes \cite{wollman2005}. An alternative mechanism proposed to resolve this discrepancy is the nucleation of MTs directly from kinetochores \cite{oconnell2007}, which was incorporated in a computational model treating chromosomal movement as random fluctuations in three dimensions \cite{paul2009}.

Describing  motor driven chromosome dynamics and MT attachment \cite{hill1985,bertalan2014} has also been the object of several computational studies mainly focusing on chromosome oscillations \cite{joglekar2002,civelekoglu2006}. These one-dimensional models do not account for congression, because they do not consider peripheral chromosomes, not lying between the spindle poles at NEB, which are, however, experimentally observed in mammalian cells \cite{barisic2014} Three dimensional numerical models have been extensively introduced to study cell division in yeast \cite{gardner2005,chacon2013,gluncic2015} but in that case motor proteins are not essential for congression and there is no NEB. It is not therefore not clear to which extent these models can be applied to mammalian cells. 

Despite the number of insightful experimental and theoretical results, it is still unclear how a collection of deterministic active motor forces interact with a multitude of randomly changing MTs to drive a reliable and coherent congression process in  a relatively short time. A key factor that has been completely overlooked in previous studies is the role of the number of MTs  composing the spindle. This is because, on the one hand, it is very difficult to measure this number experimentally in a dividing cell: The only measurement to our knowledge is reported in an early paper estimating the number of MTs in the mitotic spindle of kangaroo-rat kidney (PtK) cells as larger than $10^4$ \cite{mcintosh1975}. On the other hand, computational
limitations have restricted the number of simulated MTs to justs  few hundred \cite{magidson2011,wollman2005,paul2009}.
Yet the misregulation of several biochemical factors controlling MT nucleation (e.g.  the centrosomal protein 4.1-associated protein CPAP \cite{hung2004}) or MT depolymerization (e.g. the mitotic centromere-associated kinase or kinesin family member 2C MCAK/Kif2C \cite{bakhoum2009,stout2006,domnitz2012}) are known to affect congression, suggesting
that the number of MTs should indeed play an important, but as yet unexplored, role in the process. 

Here we tackle this issue by introducing a three dimensional model of motor driven chromosome congression and bi-orientation during mitosis involving a large number of randomly evolving MTs. Our model describes accurately the processes of stochastic search-and-capture by MTs and deterministic motor-driven transport, reproducing accurately experimental observations obtained when individual motor proteins were knocked down\cite{antonio2000,kapoor2006,putkey2002,silk2013,barisic2014}. Furthermore, the model allows us to explore ground that is extremely difficult to cover experimentally and vividly demonstrate the crucial role played by the number of MTs to achieve successful chromosome congression and bi-orientation. Increasing the number of MTs enhances the probability of bi-orientation but slows down congression of  peripheral chromosomes due to the increase of PEFs with the number of MTs. Conversely when the number of MTs is too low, congression probability is increased but bi-orientation is impaired. Most importantly, the numerical value of the optimal number of MTs is around $10^4$, which agrees with experimental estimates \cite{mcintosh1975}
but is two orders of magnitude larger than the numbers employed in previous computational studies
\cite{magidson2011,wollman2005,paul2009}.

\section*{Materials and Methods \label{sec:model}}

We consider a three-dimensional model for chromosome congression and bi-orientation in mammalian cells based on the coordinated action of three motor proteins and a large number of MTs emanating from two spindle poles.  Chromosomes and MTs follow a combination of deterministic and stochastic rules. Attached chromosomes obey a deterministic overdamped equation driven by motor forces and use MTs as rails, but attachments and detachments occur stochastically. Similarly, MTs grow at constant velocity but can randomly switch between growing and shrinking phases. The dynamics is confined within the cell cortex, modelled as a hard envelope that repels MTs and chromosomes. We set the cortex major principal axis $a$ parallel to the $x$ axis, and the minor axes as $b=0.9a$ and $c=0.7a$ parallel to the $y$ and $z$ axes, respectively. This results in a slightly flattened but almost circular cell. $n_C=46$ chromosomes are initially uniformly distributed in a sphere of radius $0.65a$ representing the nuclear envelope.

\subsection*{Microtubules}

We assume that spindle poles are already separated and kept at a constant distance throughout the congression/bi-orientation process \cite{sharp-rev-2000}, in positions $(\pm a/2,0,0)$. MTs  emanate from each pole radially as straight lines in random spatial directions. A fraction $p_{\textrm{sc}}$ of interpolar MTs forms a stable scaffold, and the remainder grow or shrink with velocities $v_g$ and $v_s$, following the dynamical instability paradigm \cite{mitchison-kirschner-1984}. In this paradigm, the transition from growing to shrinking, known as catastrophe, occurs with rate $p_{\textrm{cat}}$ and the reverse process, known as rescue, occurs with rate $p_{\textrm{res}}$. Following Ref. \cite{akiyoshi2010}, the rate of MT catastrophe and rescue both depend on the force $F$ acting on the tip of the MT as $p_{\textrm{cat}}=p_{\textrm{cat}}^0 \exp (-F/F_{\textrm{cat}})$ and  $p_{\textrm{res}}=p_{\textrm{res}}^0 \exp (F/F_{\textrm{res}})$, where $F_{\textrm{cat}}$ and $F_{\textrm{res}}$ are the sensitivities of the processes. In our simulations, the only forces on the MTs are due to end-on attachments with kinetochores, which we describe in detail below.
In most simulations, we consider a constant number of $N_{\textrm{MT}}$, but we also study the case of in which MTs nucleate at rate $k_{\textrm{nucl}}$ from each pole.

\subsection*{Chromosomes}

\begin{figure}[p]
	\begin{center}
		\includegraphics[width=0.5\columnwidth]{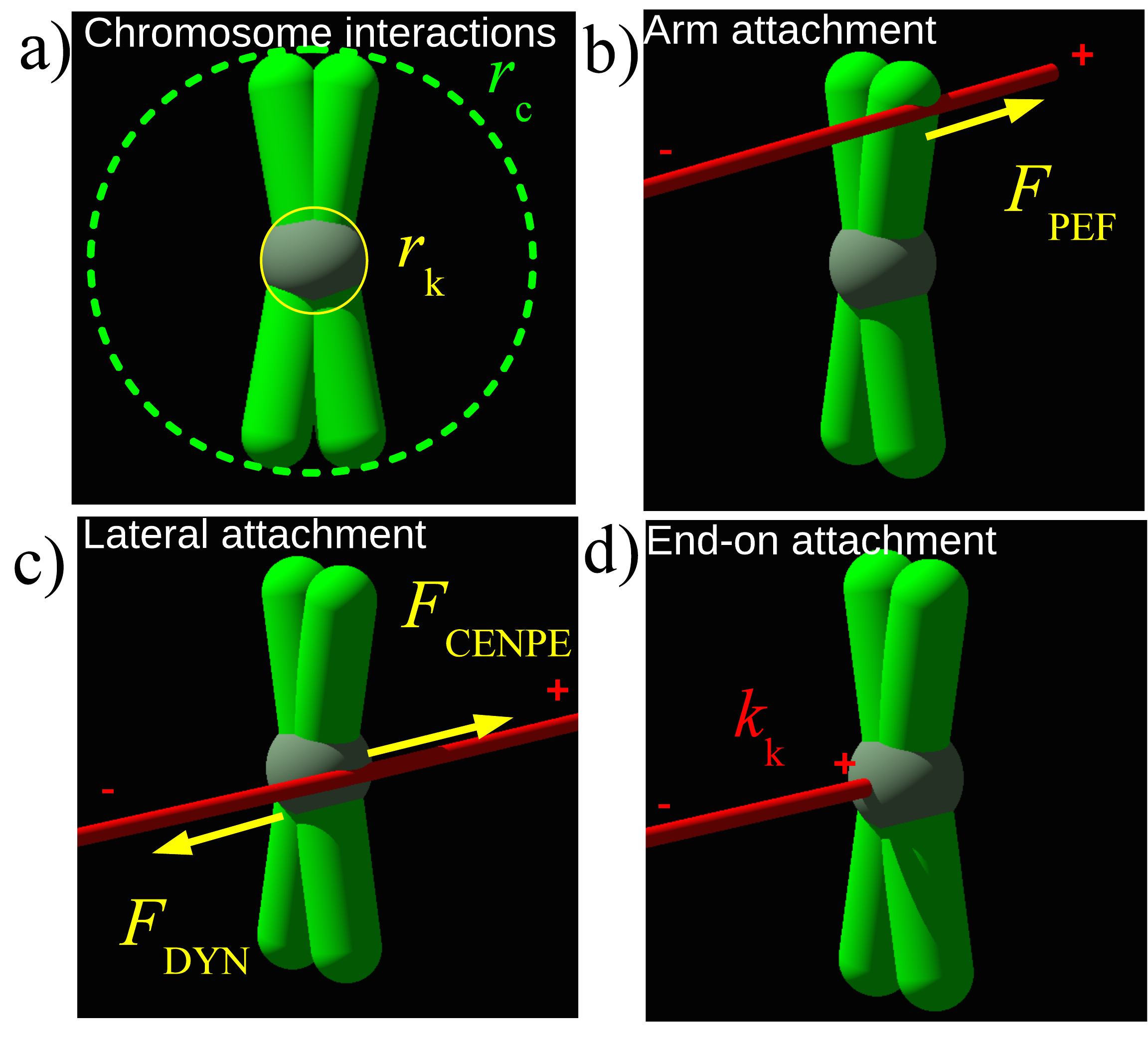}
	\end{center}
	\caption{
		Schematic of the chromosome model and forces acting on it. a) The chromosome consists of freely rotating arms 
		and of a sphere of radius $r_{\textrm{k}}$, representing the kinetochore. In the model the arms is represented by
		a disk of radius $r_{\textrm{C}}$, corresponding to the chromosome cross-section, and the kinetochore by a sphere
		of  radius $r_{\textrm{k}}$.  Microtubules (red) interact with the chromosome and exert forces on it. b) A MT passing through a chromosome arm, adds a force $F_{\textrm{PEF}}$ in the direction of the plus-end of the MT. c) Lateral attachments add constant forces originating from groups of motor proteins at the kinetochore. Which group, dynein or CENP-E is active, is determined by the simulation and described in detail in the main body of the text. d) MT tips can form end-on attachments with the kinetochore, which is represented by a harmonic spring with stiffness $k_{\textrm{k}}$ and zero rest length.}
	\label{fig:cartoon}
\end{figure}

Chromosomes consist of two large cylindrical objects, the chromatid sisters, joined at approximately their centers. Chromosome arms are floppy, with an elastic modulus around 500 Pa \cite{nicklas1983, marshall2001, marko2003} but they tend to be aligned on a plane by PEFs \cite{barisic2012}. We therefore treat chromosome arms as a two dimensional disk
of radius $r_{\textrm{C}}$, representing the cross-section for their interaction with MTs (see Fig \ref{fig:cartoon}a). 
At the centre of each chromosome sit two kinetochores, highly intricate protein complexes fulfilling a wide variety of tasks, chief of which is interacting with MTs. In the model, the two kinetochores are treated as a sphere of radius $r_{\textrm{k}}$ 
defining the interaction range with MTs (see Fig \ref{fig:cartoon}a).

Chromosomes can interact with MTs in three distinct ways: PEFs (Fig \ref{fig:cartoon}b), lateral attachments (Fig \ref{fig:cartoon}c) and end-on attachments (Fig \ref{fig:cartoon}d). Each of these interactions is associated with a specific motor force, as illustrated in the schematic in Fig \ref{fig:cartoon} and described below. 

Time is discretized and at each time step $\Delta t$ we first implement stochastic events in parallel, then perform MT growth/shrinking and update chromosome positions $\mathbf{r}_i$ according to the discretized overdamped equations of motion 
\begin{equation}
\mathbf{r}_i(t+\Delta t)=\mathbf{r}_i(t)+ \mathbf{F}_{i} \Delta t /\eta
\end{equation}
where $\eta$ is the drag coefficient and $\mathbf{F}_{i}$ is the total motor force acting on chromosome $i$.
The total force is the sum of PEFs, $F_{\textrm{PEF}}$, lateral attachment forces
due to dynein, $F_{\textrm{dynein}}$ and CENP-E, $F_{\textrm{CENPE}}$, and end-on-attachment spring forces $F_{\textrm{k}}$. The precise form of these forces is described in detail below. 

\subsubsection*{Polar ejection forces} 

For every MT crossing the chromosome within a distance $r_{\textrm{C}}$ of its geometrical center (Fig \ref{fig:cartoon}b), the chromosome acquires a PEF $F_{\textrm{PEF}}$ due to motors sitting at the chromosome arms \cite{marshall2001}, in direction of the plus end of the MT. 

\subsubsection*{Lateral attachments}

In our model, lateral kinetochore-MT attachments form when a MT crosses the kinetochore interaction sphere of radius $r_{\textrm{k}}$. Then the MT serves as a track along which the chromosome is slid by one of two groups of motor proteins, CENP-E or dynein. CENP-E applies a force $F_{\mathrm{CENPE}}$ towards the plus end of the MT, away from the spindle pole, while dynein applies a force $F_{\mathrm{DYN}}$ towards the minus end of the MT, thus pointing in the direction of the spindle pole, as illustrated in Fig \ref{fig:cartoon}c. Since we use overdamped dynamics, a constant force corresponds to a constant velocity with which the group of motor proteins moves the chromosome. To determine which type of motor is active, we take a deterministic approach motivated by  experimental results \cite{barisic2014}: we initially set CENP-E as the active motor for chromosomes that are inside a shell of radius $0.45a$ and dynein for the rest of peripheral chromosomes. Experiments show that dynein brings peripheral chromosomes to the poles \cite{rieder1990,li2007,yang2007,vorozhko2008} and is then inactivated by the action of the kinase Aurora A, while CENP-E  is activated \cite{kim2010}. We simulate this by switching off dynein at the pole and replacing it by CENP-E.

The CENP-E motor prefers to walk on long-lived MTs \cite{kim2010}, giving the chromosome a necessary bias to congress at the cell center. The biochemical factor underlying this process has been recently identified with the detyrosination of spindle microtubules pointing towards center of the cell \cite{barisic2015}. In the model, we form lateral attachments when 
CENP-E is active only if the MT has a lifetime larger than $\tau_{MT}=60s$.

\subsubsection*{End-on attachments}
\label{endon}
The two kinetochores in our model are represented as half-spheres and each has $N_{\textrm{k}}$ slots for end-on attachments with MTs. In general, when the tip of an itinerant MT is within distance $r_{\textrm{k}}$ of a kinetochore with available slots, the MT and the kinetochore form an end-on attachment. However, after NEB the kinetochores of peripheral chromosomes are covered by dynein, inhibiting end-on attachments \cite{vorozhko2008}. Hence, we allow for end-on attachments only when CENP-E is active. 
The force on the chromosome from an end-on attached MTs is translated via a harmonic coupling with zero rest length and spring constant $k_{\textrm{k}}$. 

MTs can detach stochastically from kinetochores with a rate that depends on the applied force and on the
stability of the attachment\cite{maresca-salmon2010}. Biochemical factors, such as Aurora B kinase, ensure that faulty attachments are de-stabilized  \cite{cimini2006,lampson-cheeseman-2011} and correct attachments stabilized. In particular, intra-kinetochore tension in bi-oriented chromosomes
inhibits the de-stabilizing effect of Aurora B kinase on end-on attachments \cite{lampson-cheeseman-2011}. Furthermore,
stabilization of chromosomes at the central plate is also due to action of kinesin-8 motors \cite{stumpff2008,stumpff2011,stumpff2012}.
In the present model, we simply  stabilize attachments if both kinetochores have end-on attached MTs stemming from both poles, while we treat as unstable the cases in which only a single kinetochore has end-on attachments or in which two kinetochores have end-on attached MTs all stemming from a single pole.

Unstable attachment detach with a probability that decreases exponentially with applied force 
$p^{(u)}_{\textrm{detach}}=p^{(u),0}_{\textrm{detach}}\exp(F/F_{\textrm{detach}}^{(u)}) $, where $F$ is the force on the MT tip due to coupling with the kinetochore and $F_{\textrm{detach}}^{(u)}$ is the sensitivity \cite{akiyoshi2010}. 
When the attachment is stable, we assume that the growth/shrinkage velocity of the attached MTs is slowed exponentially (see Table \ref{tab:constants} and Ref.~\cite{akiyoshi2010}), and that attachment is -- contrary to intuition -- stabilized by an applied load $p^{(s)}_{\textrm{detach}}= p^{(s),0}_{\textrm{detach}}\exp(-F/F_{\textrm{detach}}^{(s)})$. This peculiar behavior, known as a {\it catch-bond}, has been revealed experimentally \cite{akiyoshi2010} and explained theoretically \cite{bertalan2014}.

\subsection*{Implementation}
The numerical solution is implemented in a custom made C++ code. Images and videos are rendered in 3D using Povray. Simulation and rendering codes are available at 
\url{https://github.com/ComplexityBiosystems/chromosome-congression}
All parameters used in the model are summarized in Table \ref{tab:constants}. Where experimentally-measured parameters are not available, we have used estimated values. We have tested these to ensure simulation results are robust against changes in parameter values.

\begin{table*}[!ht]
\caption{\bf{Model Parameters}}
\begin{tabular}{|l|c|c|l|}
\hline
Name & Symbol &Values used &Comment/Reference \\
\hline
Cell major axis & $a$ & 15$\mu$m & estimate \\ 
Effective kinetochore radius & $r_{\textrm{k}}$ & 0.3 $\mu$m & estimate \\
Kinetochore slots & $N_{\textrm{k}}$ & 25-50 & based on PtK1 cells\cite{mcdonald1992}\\
Kinetochore--MT spring &$k_{\textrm{k}}$ & 100.0 pN/$\mu$m & magnitude similar to \cite{civelekoglu2006,joglekar2002}\\
Unstable detach rate & $p^{(u),0}_{\textrm{detach}}$ & $0.1$/s & estimate, unloaded \cite{akiyoshi2010} \\
Unstable detach sensitivity & $F^{(u)}_{\textrm{detach}}$ & 4 pN & estimate \\
Stable detach rate & $p^{(s),0}_{\textrm{detach}}$ & 0.001/s & estimate, unloaded \\
Stable detach sensitivity & $F^{(s)}_{\textrm{detach}}$ & 4 pN & estimate \\
Chromatid radius & $r_{\textrm{C}}$ & 1.1-1.5 $ \mu $m& estimate \cite{salmon1984}\\
Number of chromosomes & $n_C$ & 46 & human cell\\
PEF & $F_{\textrm{PEF}}$ & 0.5 pN& per MT \cite{brouhard2005}\\
CENP-E force & $F_{\textrm{CENPE}}$ & 5$\times$10 pN & total group\\
&&& based on stall force \cite{kim2008} \\
Dynein force & $F_{\textrm{DYN}}$ & 1.0$\times 50$ pN & per group\\
&&&based on stall force \cite{mallik2004} \\
\hline
MT growth velocity & $v_{\textrm{g}}$ &12$\mu$m/min & \cite{rusan2001}, unloaded\\
MT growth sensitivity & $F_{\textrm{g}}$ &6pN &  \cite{akiyoshi2010}\\
MT shrinking velocity & $v_{\textrm{s}}$&14$\mu$m/min & \cite{rusan2001}, unloaded \\
MT shrinking sensitivity & $F_{\textrm{s}}$ &4pN &  \cite{akiyoshi2010}\\
Rescue rate & $p^0_{\textrm{res}}$& 0.045/s & \cite{rusan2001}, unloaded\\
Rescue sensitivity & $F_{\textrm{res}}$ &2.3pN &  \cite{akiyoshi2010}\\
Catastrophe rate & $p^0_{\textrm{cat}}$ &0.058 - 0.58/s & \cite{rusan2001}, unloaded and overexpression\\
Catastrophe sensitivity & $F_{\textrm{cat}}$ &2.4pN &  \cite{akiyoshi2010}\\
Tot. number of MTs & $N_{\textrm{MT}}$  & 900 -- 30000  & \\
Fraction of linked MTs & $p_{\textrm{sc}}$  &0.1 & estimate\\

\hline
Drag coefficient & $\eta$ & $10^{-7}$ Kg/s & estimate based \\
& & & on cytoplasmic viscosity\cite{salmon1984}\\
\hline
\end{tabular}
\begin{flushleft}List of parameter values employed in the simulations.
\end{flushleft}
\label{tab:constants}
\end{table*}

\section*{Results}

\subsection*{Control of MT number by MT nucleation rate}
\begin{figure}[p]
	\begin{center}
		\includegraphics[width=0.6\columnwidth]{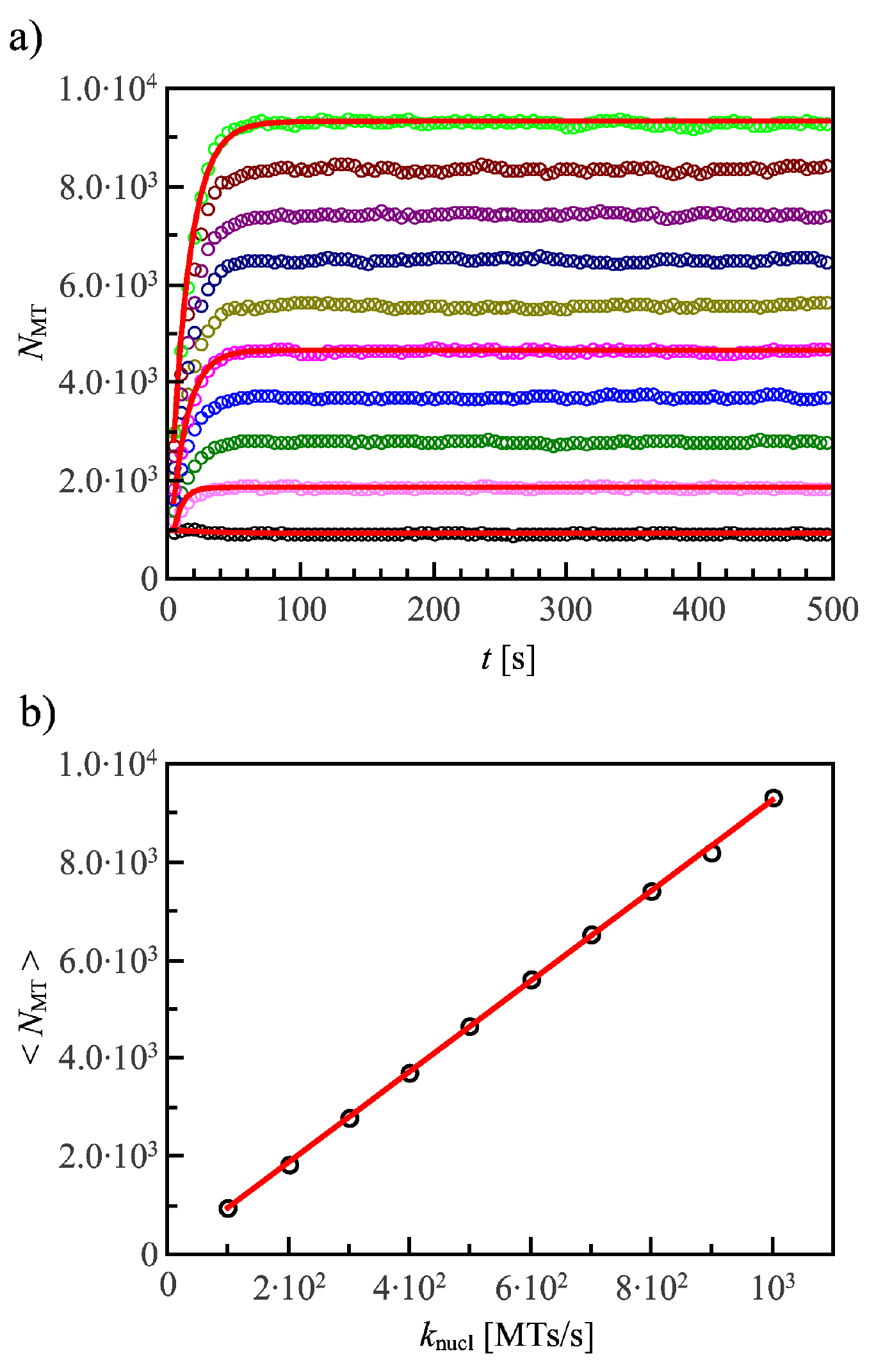}
	\end{center}
	\caption{
		The rate of microtubule nucleation controls their number.  (a) The number of MTs reaches a constant value in a time that is much shorter than the typical congression time. Different curves refer to different values of $k_{\textrm{nucl}}$. (b) The number of MTs is proportional to the rate of nucleation $k_{\textrm{nucl}}$. The numerical results here refer to a single pole. Lines are fits with the theory discussed in the text. The curves have been obtained by averaging over
		$n=1000$ independent runs of the simulations. Error bars are smaller than the plotted symbols.}
	\label{fig:Figure1}
\end{figure}

In most of our simulations, the number of MTs is fixed. To justify this, we have performed simulations in which MTs nucleate from the two spindle poles with a rate $k_{\textrm{nucl}}$. At the beginning of the simulation, we assume that the mitotic spindle is already formed, the nuclear envelope is broken, and 46 chromosomes are randomly distributed in a spherical region enclosing the poles. We then integrate the equations of motion for each chromosome and monitor the number of MTs $N_{\textrm{MT}}$ as a function of the nucleation rate $k_{\textrm{nucl}}$. We find that after a transition time (approximately 50s), that is much shorter than the congression time (Fig \ref{fig:Figure1}a),   the number of MTs fluctuates around a constant value $\langle N_{\textrm{MT}} \rangle$ that is linearly dependent on $k_{\textrm{nucl}}$ (Fig \ref{fig:Figure1}b).

The result shown in Fig \ref{fig:Figure1} can be understood from a simple kinetic equation for the number of MTs 
\begin{equation}
\frac{dN_{\textrm{MT}}}{dt} = k_{\textrm{nucl}}-k_{\textrm{out}} N_{\textrm{MT}},
\label{eq:MTkinetics}
\end{equation}
 where the second term on the right-hand side is the total rate of MT collapse.
 The rate of collapse per MT, $k_{\textrm{out}}$, is the inverse of the MT lifetime, proportional
to the MT half-life.  The solution of  Eq. \ref{eq:MTkinetics} 
 \begin{equation}
 N_{\textrm{MT}} = k_{\textrm{nucl}}/k_{\textrm{out}} (1-\exp (- k_{\textrm{out}} t) ) 
 \label{eq:solution}
 \end{equation}
 provides an excellent fit to the data with $k_{\textrm{out}}=0.09 {\textrm{s}}^{-1}$ (Fig \ref{fig:Figure1}a). The theory also shows that for long times, $t \gg 1/k_{\textrm{out}}$,  the number of MTs approaches $ N_{\textrm{MT}} = k_{\textrm{nucl}}/k_{\textrm{out}}$. Hence the number of MTs is essentially constant during the congression process, 
depending only on the rate of nucleation and collapse, which are controlled by several biochemical factors. Based on this result, we ignore the transient and keep $N_{\textrm{MT}}$ constant during each simulation.

\subsection*{Incorrect chromosome congression due to knock-down of motor proteins}
After nuclear envelope breakdown, there are two possible scenarios for congression. In the first
case, all chromosomes already lie between the poles, and have access to stable MTs. Hence,  CENP-E overcomes dynein, moving the chromosome directly towards the center of the cell. The second scenario involves chromosomes not having access to stable MTs, because their initial position does not lie between the poles. Those chromosomes are first driven by dynein to the nearest pole
and remain there until they find a stable MT to which they attach laterally. At this point, they  slide  towards the central plate using CENP-E motor on the stable MT. We show the evolution of these two scenarios in  S1 and S2 Videos. In all simulations we ran with the present parameters ($n>30$ instances per scenario) all chromosomes congress and bi-orient.

\begin{figure*}[p]
	\begin{center}
		\includegraphics[width=0.8\columnwidth]{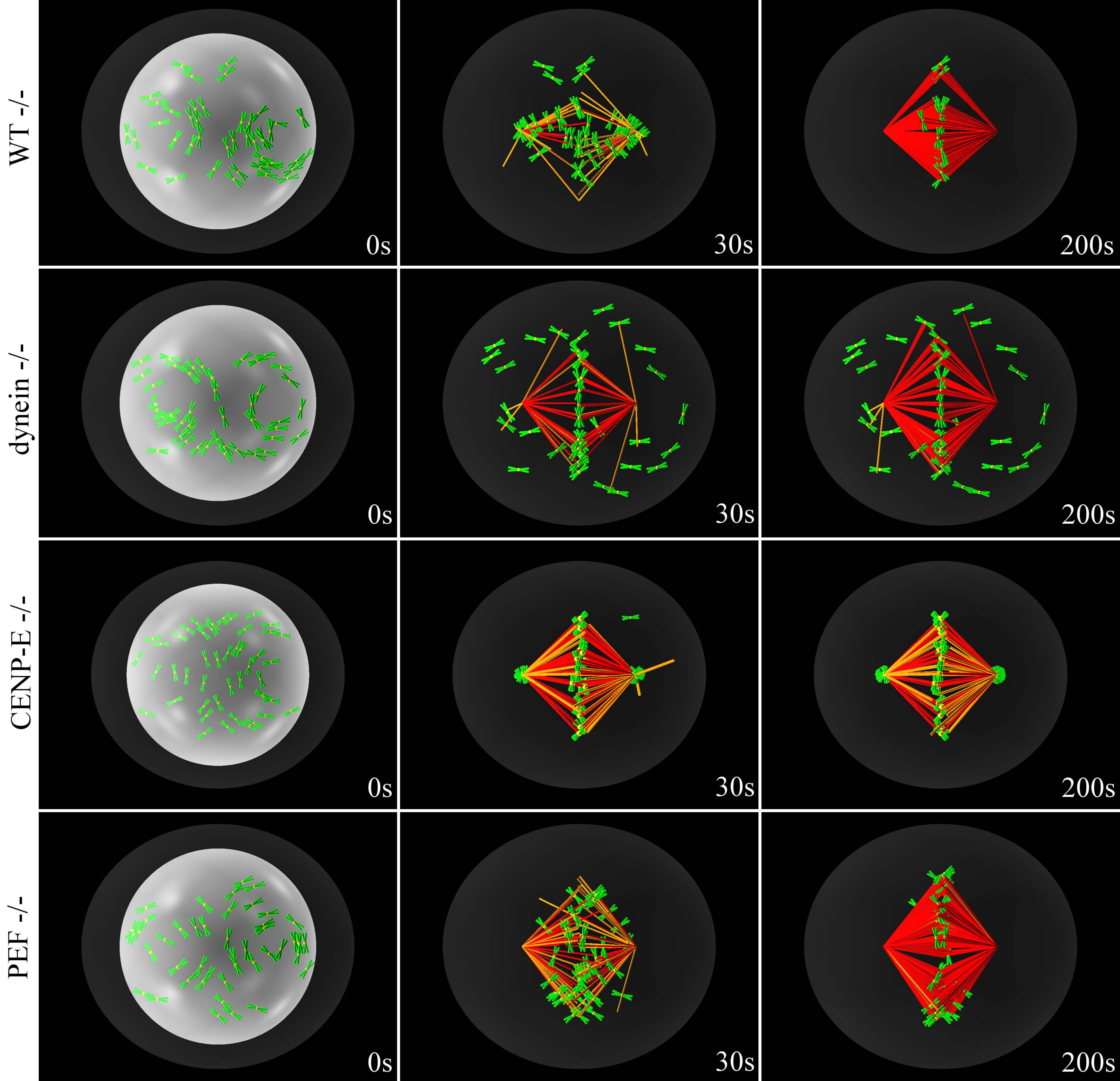}
	\end{center}
	\caption{Time-lapse snapshots of the simulated congression process when motors are suppressed. Chromosomes are shown as having chromatid arms (green) for viewing purposes, while the kinetochores are shown as yellow spheres. Not all MTs are shown, only those that serve as rails for kinetochore motor-proteins (orange) and end-on attached MTs (red). The nuclear envelope is shown for reference in each of the first panels as a white sphere. The cortex is represented in dark grey. The wild type (WT) case, in which all motor proteins are active, is shown for comparison in row 1. When dynein is suppressed (row 2),  PEFs push peripheral chromosomes to the cortex. However, when all chromosomes start between the poles, congression takes place normally.
		When CENP-E is depleted (row 3), peripheral chromosomes or other chromosomes that are transported to the poles get trapped there.  Depleting PEFs (row 4) delays congression significantly and destabilizes the coherence of the central plate. It makes no difference whether chromosomes start all between poles or there are peripheral chromosomes. 
	}
	\label{fig:Figure2}
\end{figure*}

Next, we switch off motor proteins individually (dynein, CENP-E or PEF) to show that the model successfully reproduces what happens in cells, where all these motors are essential. The
results are summarized in Fig~\ref{fig:Figure2} (see also S3,S4 and S5 Videos) and show that the suppression of each of the motors leads to incorrect congression or bi-orientation. Suppressing kinetochore dynein does not allow peripheral chromosomes to congress, as shown in row 2 of Fig~\ref{fig:Figure2}. Deletion of CENP-E traps chromosomes at the poles, as shown in row 3, and PEF knockdown severely reduces  the cohesion of the central plate where chromosomes can not bi-orient, as shown in row 4.

These knock-downs have also been studied experimentally, yielding results in line with ours.
In Refs.~\cite{antonio2000,kapoor2006} principal contributors to PEF are knocked down, and it is shown that in cases where there are no peripheral chromosomes, the chromosomes can congress but are not stable at the central plate. Furthermore, other experiments show that chromosomes are also stabilized at the central plate due to the effect of the kinesin-8 Kif18A on MT plus ends \cite{stumpff2008,stumpff2011,stumpff2012}. These observations fold neatly into our model and yield a possible explanation of the above mentioned slowing down of MT plus ends at kinetochores. It should also be noted that when the effect of Kif18A is removed and MT plus ends follow fast dynamics again, the effective PEFs in the vicinity of the central plate are reduced, further destabilizing chromosome alignment \cite{barisic2014}.
In Refs.~\cite{putkey2002,silk2013}, on the other hand, CENP-E is knocked down or suppressed, and results in chromosomes being trapped at spindle poles. Finally. in Ref.~\cite{barisic2014} all three motors are suppressed individually, with exactly the same results as presented here from our simulations. 

\subsection*{Optimal number of MTs for chromosome congression and bi-orientation}

\begin{figure*}[p]
	\begin{center}
		\includegraphics[width=0.9\columnwidth]{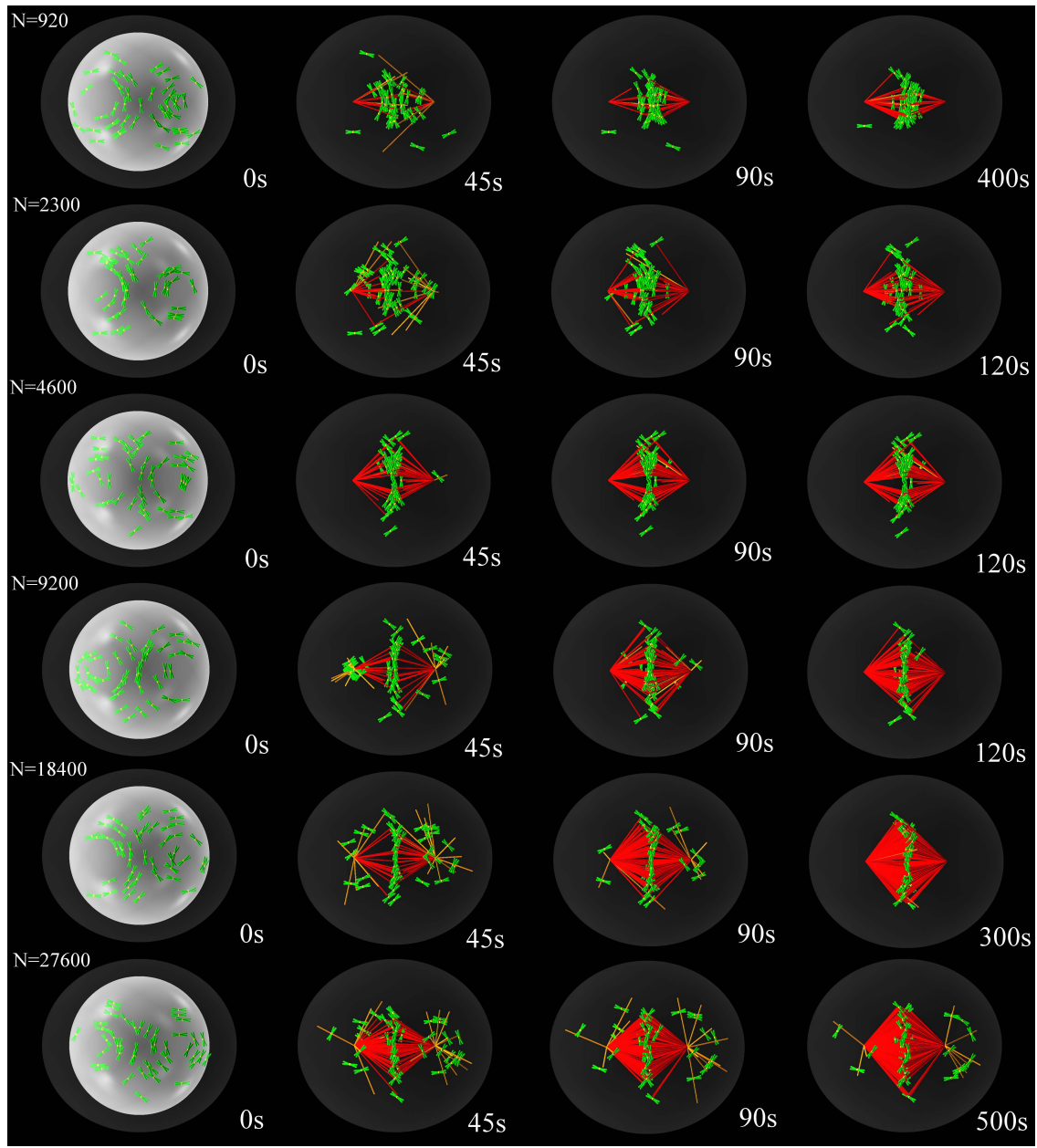}
	\end{center}
	\caption{
		Time-lapse snapshots of the simulated congression process for different values of the number of MTs per kinetochore. Congression fails if this number is too small or too large. Chromosomes are shown as having chromatid arms (green) for viewing purposes, while the kinetochores are shown as yellow spheres. Not all MTs are shown, only those that serve as rails for kinetochore motor-proteins (orange) and end-on attached MTs (red). The nuclear envelope is shown for reference in each of the first panels as a white sphere. The cortex is represented in dark grey.
	}
	\label{fig:Figure3}
\end{figure*}

\begin{figure}[p]
	\begin{center}
		\includegraphics[width=0.35\columnwidth]{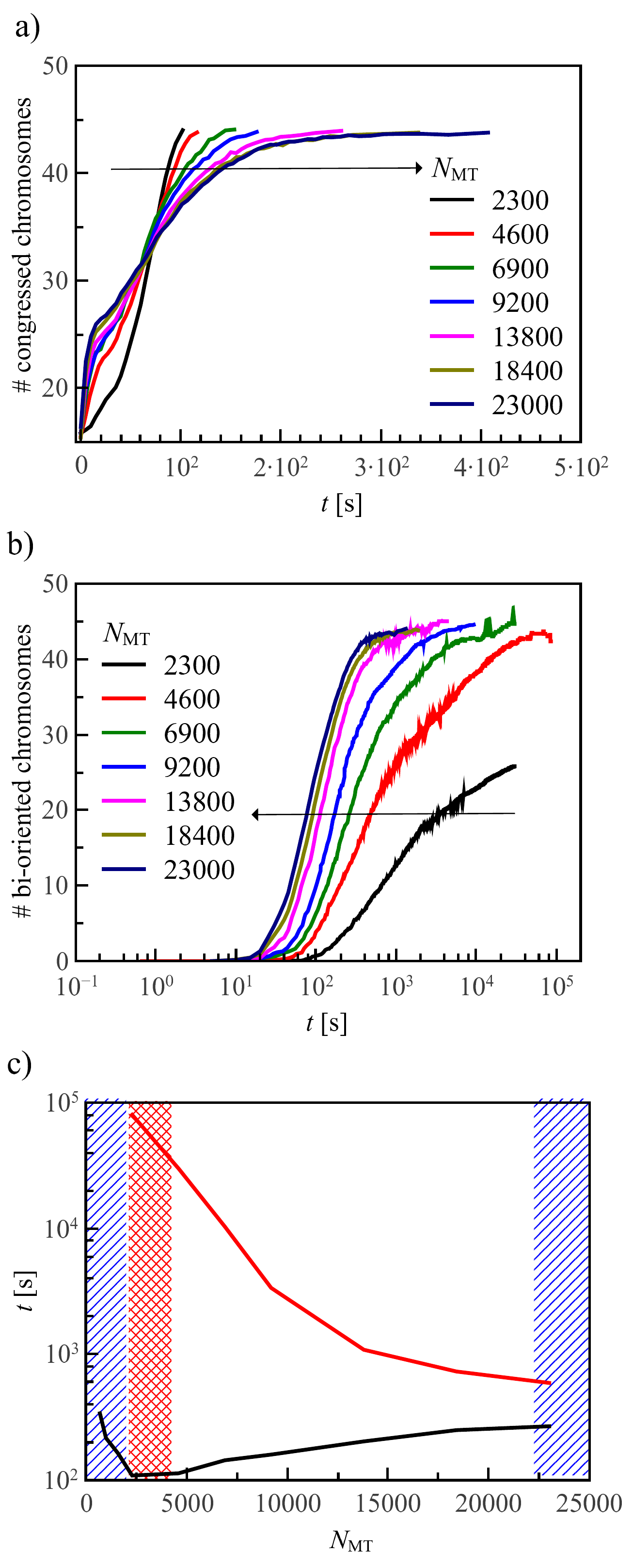}
	\end{center}
	\caption{
		The distribution of (a) congression and (b) bi-orientation times for various MT densities. The arrows indicate the trends for increasing MT densities. Congression is faster for a lower number of MTs per kinetochore, because PEFs are directly proportional to the number of MTs. However, bi-orientation is much slower for low MT densities, because the time needed to find every kinetochore is strongly influenced by the number of MTs. This is summarized in (c) showing  the time $t_{p=1/2}$ for which the congression/bi-orientation probability is one half. The maximum waiting time for congression is $10^3$s and for bi-orientation $10^5$s. If the MT density is too low, not all samples bi-orient, as indicated by the red shaded area. Decreasing the MT density even further severely reduces the congression probability, indicated by the blue shaded area. On the other hand, increasing the MT density too much also impairs congression since kinetochore dynein will not be strong enough to overcome PEFs. These results show that there is a sweet spot for congression/bi-orientation as a function of the number of MT, lying between $7\times10^3$ and $1.8\times10^4$ MTs. All curves have been obtained by averaging over
		$n=100$ independent runs of the simulations. Error bars are smaller than the plotted curves.
	}
	\label{fig:Figure4}
\end{figure}

We find that the ratio of the number of total MTs in the system divided by twice the total number of chromosomes that is,
the total number of kinetochores, affects the congression process in a non-trivial manner, as illustrated in Fig \ref{fig:Figure3} and S6 and S7 Videos. In particular, chromosome congression and bi-orientation are influenced by the number of MT in opposite ways: While a large number of MTs enhances the chances of bi-orientation, it slows down congression. 
This is due to the fact that PEFs increase with the number of MTs, thus acting against kinetochore dynein and possibly hindering the motion of peripheral chromosome towards the poles. In the wild-type case, kinetochore dynein in usually strong enough to overcome these PEFs \cite{barisic2014}. Overexpression of motors giving rise to PEFs can have adverse effects, such as the over-stabilization of kinetochore-MT attachments \cite{cane2013}. On the other hand, stabilizing MTs by disrupting various MT-depolymerase chains results in much slowed down congression and bi-orientation \cite{tanenbaum2011}.
We show the effect of too strong PEFs on our model in Fig \ref{fig:Figure4}a, where the distribution of congressed chromosomes is plotted versus time for different MT densities. On the other hand, PEFs stabilize congressed chromosomes at the central plate, and in a simple search and capture scenario \cite{holy-leiber-1995}, like the one implemented in our model, the more MTs there are the faster chromosomes become bi-oriented, as indicated in Fig \ref{fig:Figure4}b. 
In Fig \ref{fig:Figure4}c, we plot the median of the congression/bi-orientation time distribution defined as the time for which the probability of congression (black) and bi-orientation (red) is one half. At very low MT densities (blue shaded area), reported in the left-hand-side of Fig  \ref{fig:Figure4}c, not all samples congress within the limit of $10^3$ seconds. At slightly higher MT densities (red shaded area), not all samples bi-orient within the limit of $10^5$ seconds. Finally, at very high MT densities,  PEFs become so strong that they reduce the congression probability. These observations indicate the existence of a {\it sweet spot} for the MT density suggesting that successful congression and bi-orientation can only happen only if the total number of MTs in the spindle lies in the range of 7$\cdot 10^3-1.8\cdot 10^4$. 

\subsection*{Overexpressing MT depolymerases reduces the congression probability}

\begin{figure}[p]
	\begin{center}
		\includegraphics[width=0.7\columnwidth]{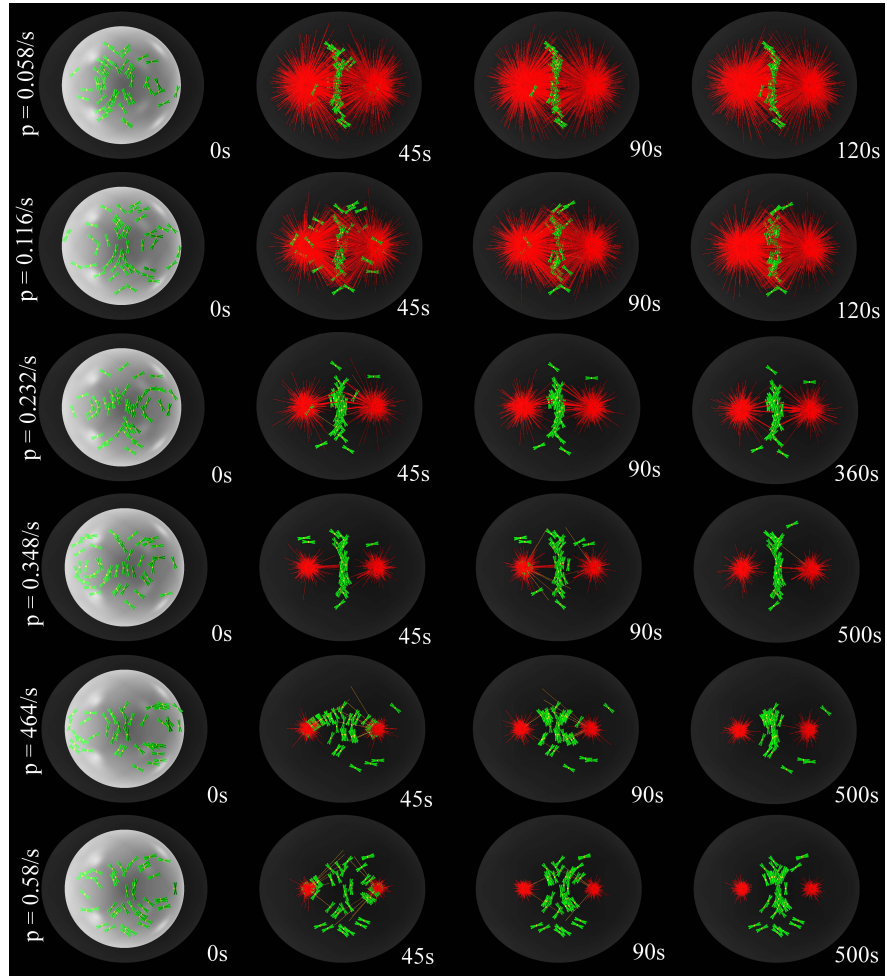}
	\end{center}
	\caption{
		Time-lapse snapshots of the simulated congression process for different values of the rate of MT catastrophes $p_{\textrm{cat}}^0$. Large values of $p_{\textrm{cat}}^0$, that is, overexpression of MT depolymerases, lead to unsuccessful congression.
		The nuclear envelope is shown for reference in each of the first panels as a white sphere. The cortex is represented in dark grey.}
	\label{fig:Figure5}
\end{figure}

\begin{figure}[p]
	\begin{center}
		\includegraphics[width=0.6\columnwidth]{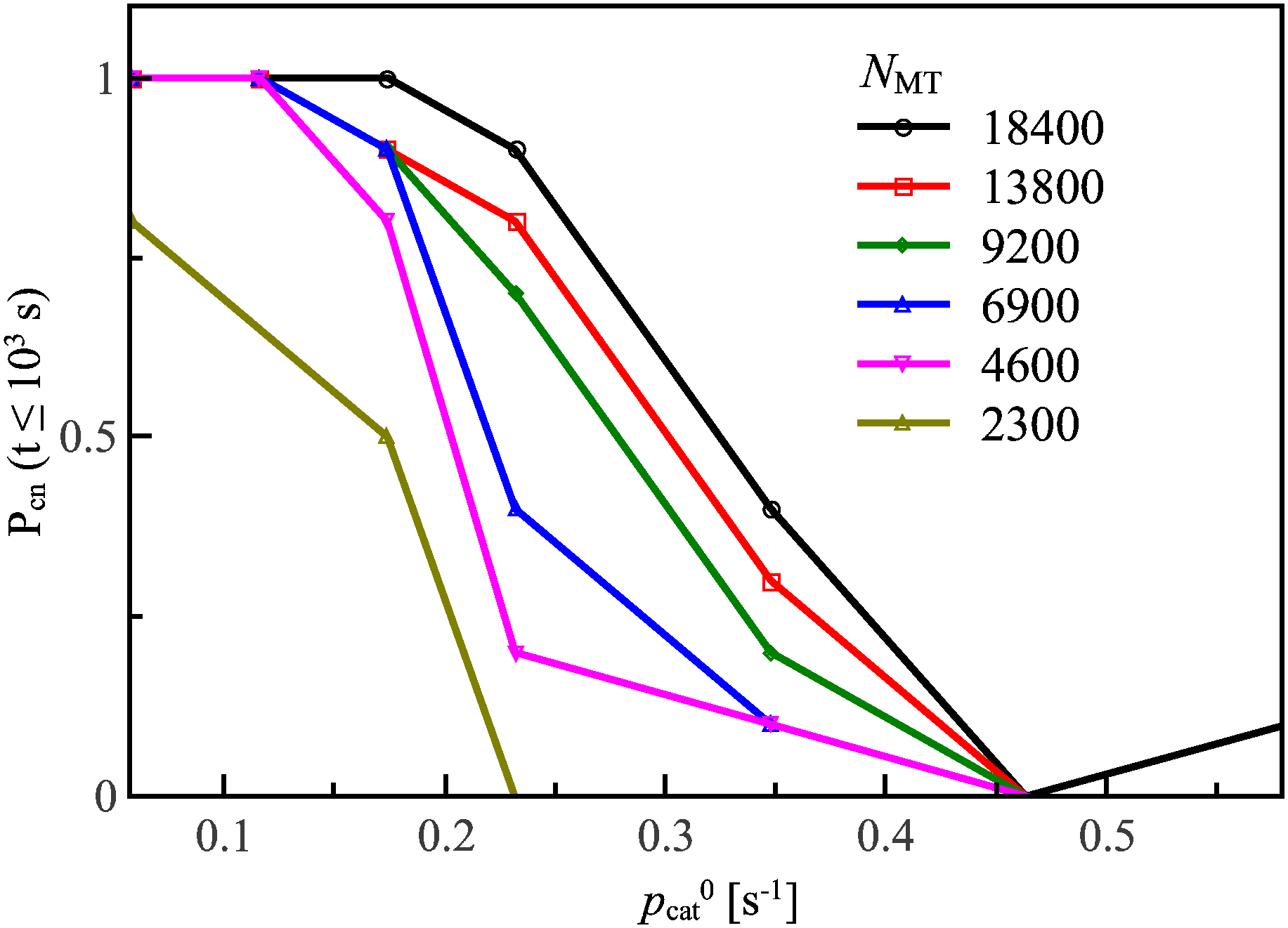}
	\end{center}
	\caption{
		Congression probability plotted against catastrophe rate.
		Overexpressing catastrophe inducing factors can severely limit the congression probability. 
		Each point represents the fraction of $n=10$ independent runs of the simulations that have reached congression during a waiting time of $10^3$s. Congression is stable over a wide range of catastrophe rates, but breaks down completely at approximately at $p_{\textrm{cat}}^0 = 0.046 \textrm{s}^{-1}$. 
	}
	\label{fig:Figure6}
\end{figure}

An experimentally testable prediction of our model is the effect on congression of the overexpression of factors 
affecting MT depolymerization \cite{ganguly2011}. The catastrophe/rescue rate ratio determines the MT length distribution during cell division. Shorter MTs would significantly hamper the search and capture process: Chromosomes lying at the extreme periphery would be harder to reach, decreasing the chances for congression. To quantify this effect, we performed $n=10$ simulations for each MT density and increasing the value of the catastrophe rate, as illustrated in Fig \ref{fig:Figure5} and in S8 Video. The corresponding congression probability is reported in  Fig \ref{fig:Figure6}. For low MT densities the effect is very drastic and even partial congression is suppressed. For the {\it sweet-spot} densities, MT depolymerases overexpression has only a small effect, until the catastrophe rate becomes too large and congression disappears.

\section*{Discussion}
Understanding cell division and its possible failures is a key problem that is relevant
for many pathological conditions including cancer. While many biochemical factors controlling several aspects of the division process have been identified, how these factors work together in a coherent fashion is still an open issue. We have introduced a comprehensive three dimensional computational model for chromosome congression in mammalian cells, using stochastic MT dynamics as well as motor-protein interplay. The model incorporates movement of the peripheral chromosomes to the poles and their escape from there towards the central plate. Contrary to previous models that only used a limited number of MTs (e.g. a few hundred in Ref. \cite{paul2009}), we are able to simulate up to $3\cdot 10^4$ MTs. McIntosh et al. reported already in 1975 that the number of MTs in the mitotic spindle of kangaroo-rat kidney (PtK) cells during metaphase is larger than $10^4$ \cite{mcintosh1975}, in good agreement with our predictions. Also, to put this number in perspective, we notice that each human chromosome has up to 50 end-on attachment slots per kinetochore, and on average 25 MTs attached \cite{mcdonald1992}. Since there are 46 chromosomes in human cells, this corresponds to 2300 attached MTs on average.  The total number of MTs in the spindle should be much larger than the number of attached MT and therefore $10^4$ MTs appears to be a reasonable number. It is interesting to remark that with this number of MTs, congression and bi-orientation of chromosomes is quick enough that the assumption of biased search \cite{wollman2005} is not needed. 

With our model we show that the total number of MTs in the spindle is {\it per se} a crucial controlling factor for successful cell division. When this number is too low or too high, congression and/or bi-orientation fail. This explains apparent paradoxes where the same factors can lead to different pathological conditions when up or down regulated. For instance, the centrosomal protein 4.1-associated protein (CPAP),  belonging to the microcephalin (MCPH) family \cite{bond2005}, is known inhibit MT nucleation \cite{hung2004}. CPAP overexpression leads  to abnormal cell division \cite{kohlmaier2009,schmidt2009}, whereas mutations in CPAP can cause autosomal recessive primary microcephaly, characterized by a marked reduction in brain size \cite{thornton2009}. In the model, we can account for CPAP overexpression by inhibiting MT nucleation, while its mutation can be simulated by increasing $k_{\textrm{nucl}}$. The two processes push the number of MTs out of its {\it sweet spot}, along different directions and therefore explain the different pathological conditions with a single
mechanism.

A similar reasoning explains the role of mitotic centromere-associated kinase or kinesin family member 2C (MCAK/Kif2C) that is localized at MT plus ends \cite{domnitz2012} and functions as a key regulator of mitotic spindle assembly and dynamics \cite{hunter2003,zhu2005} by controlling MT length \cite{domnitz2012}. Higher expression of MCAK level has been found in gastric cancer tissue \cite{nakamura2007}, colorectal and other epithelial cancers \cite{gnjatic2010} and breast cancer \cite{shimo2008}.  In fact, both depletion \cite{stout2006,bakhoum2009} and overexpression \cite{ganguly2011,tanenbaum2011} of MCAK lead to cell division errors.  From the point of view of our model, we can understand that MCAK overexpression increases the rate of MT depolymerization reducing their length and number to a level in which bi-orientation is not possible. 
Finally our model explains the recent results linking CIN to the overexpression of AURKA or the loss of CHK2, both enhancing MT assembly rate \cite{ertych2014}. Increasing MT velocity 
effectively reduces the amount of tubulin units available for MT nucleation, thus 
decreasing the number of MTs and imparing bi-orientation.

In conclusion, our model represents a general computational tool to predict the effect of biological factors on cell division making it a valid tool for {\it in silico} investigation of related pathological conditions. The main strength of our computational approach
is that can it help answer questions that are extremely difficult to address experimentally,
such as the role of the number of microtubules in driving successful cell division.


\section*{Acknowledgments}
We thank M. Barisic and H. Maiato for useful suggestions and for sharing the results of Ref. \cite{barisic2014} before
publication. We thank J. R. McIntosh for pointing out Ref. \cite{mcintosh1975} and M. Zaiser for critical reading of
the manuscript. 


\section*{Supporting Information}
%

\begin{description}

\item{\bf S1 Video. Congression of scattered chromosomes.} Representative example of the congression process in the case in which some of the chromosomes are initially scattered beyond the poles.
\item{\bf S2 Video. Congression of interpolar chromosomes.}  Representative example of the congression process in the case in which  all of the chromosomes initially lie between the poles.
\item{\bf S3 Video. Congression with PEF knockdown.} Representative example of the congression process when PEF is suppressed.
\item{\bf S4 Video. Congression with Dynein knockdown.} Representative example of the congression process when Dynein is suppressed.
\item{\bf S5 Video. Congression with CENP-E knockdown.} Representative example of the congression process when CENP-E is suppressed.
\item{\bf S6 Video. Congression with a small number of MTs.} Representative example of the congression process with 10MTs per kinetochore.
\item{\bf S7 Video. Cpngression with a large number of MTs.} Representative example of the congression process with 300MTs per kinetochore.
\item{\bf S8 Video. Congression with MT depolymerases overexpression.} Representative example of the congression process overexpressing MT depolymerases. In these simulations $p_{\textrm{cat}}^0=0.348{\textrm{s}}^{-1}$.

\end{description}

\clearpage

\end{document}